\documentclass{evn2002}

\usepackage{graphicx}
\usepackage{psfrag}
\begin{document}
   \title{Radio-Emission from Cosmic Ray Air-Showers - A Theoretical Perspective for LOPES}

   \author{T. Huege  \and  H. Falcke}

   \institute{Max-Planck-Institut f\"ur Radioastronomie, Auf dem H\"ugel 69, 53121 Bonn, Germany}

   \abstract{

High-energy cosmic ray air-showers have been known for over 30 years to emit strong radio pulses in the regime between a few and several 100~MHz (Allan \cite{Allan}). To date, however, a thorough analysis of the emission mechanisms has not yet been conducted. Adopting a simplified shower geometry and electron-positron energy distribution, we calculate theoretical pulse spectra in the scheme of synchrotron emission from highly relativistic e$^{\pm}$ pairs gyrating in the Earth's magnetic field.
These calculations will play an important role for the calibration of observational data of radio-emission from cosmic ray air-showers acquired with LOPES and later LOFAR and SKA.
             }

   \authorrunning{T. Huege and H. Falcke}
   \titlerunning{Radio-Emission from Cosmic Ray Air-Showers - Theoretical Perspective}
   \maketitle
%
%________________________________________________________________

\section{Introduction}

The pulsed radio-emission from high-energy cosmic ray air-showers allows to study their physics with forthcoming digital radio-interferometers such as LOFAR -- an approach offering a number of advantages over other methods (Falcke \& Gorham \cite{Falcke}, hereafter FG02). The aim of the LOPES project is to develop and test the necessary hardware, software and techniques for future implementation in LOFAR. Additionally, a thorough understanding of the underlying emission processes is necessary to interpret and calibrate the observational data. Past modeling efforts for radio-emission from cosmic ray air-showers have concentrated on scenarios such as charge-separation and transverse currents induced by the Earth's magnetic field (Kahn \& Lerche \cite{Kahn}). An equivalent, but conceptually more attractive and flexible approach is the scenario of coherent synchrotron emission from e$^{-}$ and e$^{+}$ gyrating in the Earth's magnetic field.

%__________________________________________________________________

\section{Emission process and model calculations}

Our emission model is based on standard synchrotron theory for highly relativistic particles as described by Jackson (\cite{Jackson}). To circumvent problems associated with retardation effects, the $\vec{E}$-field is calculated in the frequency-domain, $\vec{E}(\omega)$, rather than the time-domain, $\vec{E}(t)$. The far-field energy spectrum per unit solid angle and unit frequency ($\propto|\vec{E}(\omega)|^{2}$) generated by a single e$^{\pm}$ pair as a function of the angle of the particle trajectories to the line of sight, $\theta$, and the curvature radius $\rho$ (depending on the $\vec{B}$-field and Lorentz $\gamma$-factor) is then given by ($K_{\nu}$ = modified Bessel-function of order $\nu$):
%______________________________________________________________
   \begin{figure}
   \psfrag{nu0MHz}[c][B]{$\nu$~[MHz]}
   \psfrag{Snu0Jy}[c][t]{$S_\nu$~[Jy]}
   \includegraphics[width=8.6cm]{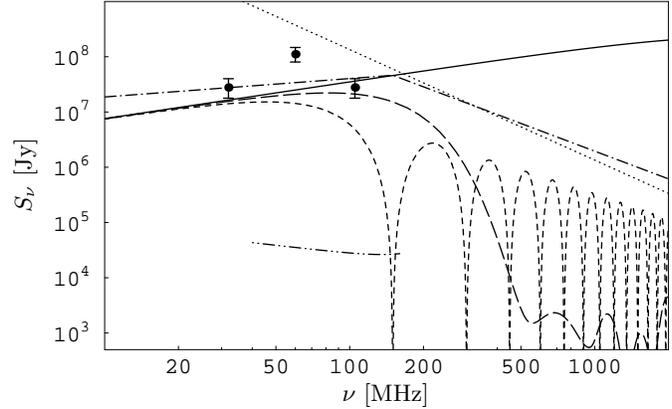}
   \caption{
   \label{fig:coherence_variants}
   Equivalent flux density spectrum at the centre of a $10^{17}$~eV shower radio pulse with $R=6$~km, $\Delta\nu=1$~MHz and $\gamma\equiv60$. Solid: full coherence, short-dashed: uniform 2~m line-charge, long-dashed: Gaussian 2~m line-charge (noise at highest frequencies being due to numerical error), dash-dotted: FG02 approximation, dotted: empirical formula (see text), dash-2x-dotted: LOPES 100 antenna station root-mean-square noise, points: Prah (\protect\cite{Prah}) data\protect\footnotemark[1] (could be subject to systematic errors due to uncertainties in primary particle energy calibration)
   }
   \end{figure}
%______________________________________________________________
%
\begin{displaymath}
\frac{\mathrm{d}^{2}I}{\mathrm{d}\omega \mathrm{d}\Omega}=\frac{4e^{2}}{3\pi c^{2}}\left(\frac{\omega \rho}{c}\right)^{2}\!\left(\frac{1}{\gamma^{2}}+\theta^2\right)^{\!2}\!\!K_{2/3}^{2}\!\!\left[\frac{\omega\rho}{3c}\left(\frac{1}{\gamma^{2}}+\theta^{2}\right)^{\!3/2}\right]
\end{displaymath}
\footnotetext[1]{Data not included in the original conference proceedings.}
which takes advantage of the symmetry arising from the pair-wise creation of e$^{-}$ and e$^{+}$. $N$ e$^{\pm}$ pairs radiating fully coherently (i.e.\ simultaneously at the same location) would yield an $N^{2}$ enhancement of this spectrum. If the e$^{\pm}$ pairs are spatially distributed, however, their contributions have to be added taking into account the appropriate phase-delays. Compared to the fully coherent case this leads to an attenuation by a coherence-factor $S(\omega)$ (see Aloiso \& Blasi (\cite{Blasi}) for an analysis of coherence effects regarding synchrotron radiation). The simplest shower geometry taking into account coherence effects is a line of length $d$ (given by the typical thickness of the air-shower ``pan-cake'' at its maximum development and set to 2~m here) on which the $N$ radiating e$^{\pm}$ pairs are uniformly distributed. A more realistic Gaussian distribution of the e$^{\pm}$ pairs along the line is also considered. The air-shower maximum consists of approximately $E_{\mathrm{p}}/$GeV particles with a mean $\gamma$ of 60 (see, e.g., Allan \cite{Allan}). The simplest approximation then is to set $\gamma\equiv60$ for all e$^{\pm}$. As a more realistic energy distribution, we adopt a broken power-law rising linearly with $\gamma$, peaking at $\gamma=60$ and declining with $\gamma^{-2}$ towards higher $\gamma$. The corresponding spectrum can then be calculated by integrating over the e$^{\pm}$ pair spectra with the normalisation such that the total energy is the same as in the monoenergetic case. Given a specific observation bandwidth, the pulse associated with the spectrum can be reconstructed as the Fourier-transform of $\vec{E}(\omega)$.
We formally define an ``equivalent flux-density'' $S_{\nu}$ as in FG02. With distance $R$ from the air-shower maximum to the observer (typically 5--7~km for a $10^{17}$~eV shower), pulse duration $\Delta t$, and taking into account bandwidth effects via the quotient of observation bandwidth $\Delta\nu$ and the (pure) synchrotron spectrum ``critical frequency'' $\nu_{\mathrm{c}}$, this is: 
\begin{displaymath}
S_{\nu}=\frac{\mathrm{d}^{2}I}{\mathrm{d}\omega \mathrm{d}\Omega}\ \frac{2\pi}{R^{2} \Delta t} \left(\frac{\Delta\nu}{\nu_{\mathrm{c}}}\right) \quad\mathrm{with}\quad \nu_{\mathrm{c}}=\frac{3e\gamma^{2}B}{4\pi m_{\mathrm{e}}c}
\end{displaymath}
%
%
%\section{Example calculations}
%
%______________________________________________________________
   \begin{figure}
   \psfrag{R0m}[c][B]{distance from shower centre [m]}
   \psfrag{Snu0Jy}[c][t]{$S_{\nu}$~[Jy]}
   \includegraphics[width=8.6cm]{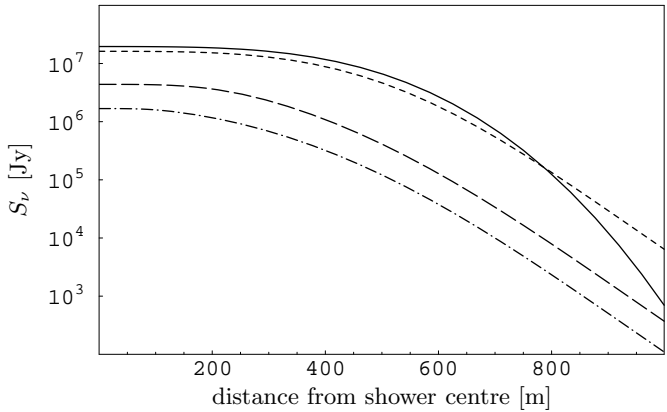}
   \caption{Radial dependence of the equivalent flux density for a $10^{17}$~eV 2~m Gaussian line-charge shower at $\nu=50$~MHz with $R=6$~km and $\Delta\nu=1$~MHz for the $\gamma\equiv60$ case (solid) and for broken power-law distributions from $\gamma=5$--120 (short-dashed), $\gamma=5$--1000 (long-dashed) and $\gamma=5$--10000 (dash-dotted).
   \label{fig:radius_dependence}}
   \end{figure}
%______________________________________________________________

Using this formalism, we can produce a set of model calculations. Fig.\ \ref{fig:coherence_variants} shows the attenuation of the fully coherent synchrotron spectrum due to interference effects. A uniform particle distribution introduces the typical $(\sin x/x)^{2}$ modulation that also appears as the interference pattern of a rectangular opening. The first minimum occurs at 150~MHz, which corresponds to $c/d$. In the Gaussian distribution, the smaller mean particle distance allows the spectrum to extend to higher frequencies but leads to stronger attenuation afterwards. Past experimental results have been described by an empirical formula (Allan \cite{Allan}). Its trend fits well with the uniform line-charge spectrum at high frequencies. As later measurements show, however, the spectrum flattens below $\sim 100$~MHz (see FG02) and the formula over-predicts the flux there. Also included is the estimated RMS noise level for a 100 antenna LOPES configuration (see FG02), which should be able to easily detect (SNR $>100$) a typical $10^{17}$~eV air-shower.
%
%For the uniform e$^{\pm}$ pair distribution, this introduces the typical $\sin(x)/x$-modulation that always appears as the interference pattern of a rectangular opening.
%______________________________________________________________
   \begin{figure}
   \psfrag{time0ns}[c][B]{t~[ns]}
   \psfrag{Efield0muV0m}[c][t]{\vec{E}(t)~[$\mu$V m$^{-1}$]}
   \includegraphics[width=8.6cm]{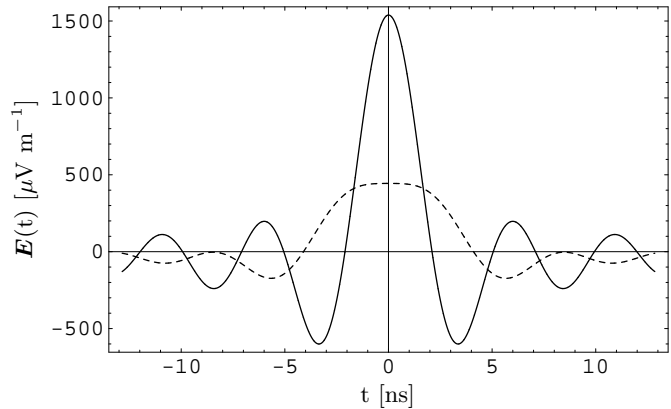}
   \caption{Reconstructed pulse in the centre of a $10^{17}$~eV shower with $\gamma\equiv60$, $R=6$~km and using an idealised rectangle filter spanning 10--200~MHz. Solid: full coherence, short-dashed: 2~m uniform line-charge
   \label{fig:pulses_0m}}
   \end{figure}
%______________________________________________________________
%
Fig.\ \ref{fig:radius_dependence} illustrates the radial dependence of the air-shower emission on the ground. One has to keep in mind, however, that a more realistic shower-geometry will distribute most of the particles away from the shower core, leading to a significant broadening as well as attenuation in the centre region. The more realistic e$^{\pm}$ energy distribution only broadens the emission pattern very slightly due to the presence of low-energy particles. The overall emission gets weaker as the total number of particles drops when more energy is distributed to high-energy e$^{\pm}$ pairs.
Fig.\ \ref{fig:pulses_0m} shows two reconstructed radio-pulses. The $\sim 6$~ns timescale of the fully-coherent pulse is given by the observation bandwidth of $\sim 200\ \mathrm{MHz}=1/5\ \mathrm{ns}$ -- the pulse is unresolved. The timescale of the uniform line-charge pulse is governed by the spectral cut-off at 150~MHz. Decreasing the bandwidth will linearly decrease the pulse amplitude, and below bandwidths of 150~MHz both pulses will become unresolved.

\section{Conclusions}

We have shown that known properties of radio-emission from extended cosmic ray air-showers can be successfully reproduced using the approach of coherent synchrotron emission from e$^{\pm}$ pairs gyrating in the Earth's magnetic field. The comparison with the FG02 approximation and empirical results is encouraging, and the next step will be to incorporate a more realistic geometry for the air-shower. All predictions show that LOPES (or a single LOFAR station) should easily be able to detect the radio pulses of a $10^{17}$~eV shower.


\begin{thebibliography}{}

  \bibitem[1971]{Allan} Allan, H.R. 1971,
      Prog. in Elem. part. and Cos. Ray Phys.,
      ed. J.G. Wilson and S.A. Wouthuysen,
      (N. Holland Publ. Co.), Vol. 10, 171

  \bibitem[2002]{Blasi} Aloiso, R., \& Blasi, P. 2002, astro-ph/0201310


  \bibitem[2002]{Falcke} Falcke, H., \& Gorham, P. 2002, Astropart. Physics in press, astro-ph/0207226


  \bibitem[1975]{Jackson} Jackson, J.D. 1975,
      Classical Electrodynamics,
      (John Wiley \& Sons, New York)

  \bibitem[1966]{Kahn} Kahn, F.D. \& Lerche, I. 1966,
      Proc. Roy. Soc., A-289, 206

  \bibitem[1971]{Prah} Prah, J.H. 1971, Radio Emission from Extensive Air Showers, MSc. Thesis, Univ. of London


\end{thebibliography}
\end{document}